\documentclass[onecolumn,preprintnumbers]{revtex4}
\usepackage{makeidx}
\usepackage{amssymb}
\usepackage{amsmath}
\usepackage{mathrsfs}
\usepackage{graphicx}
\usepackage{dcolumn}
\usepackage{bm}
\usepackage[center]{subfigure}
\usepackage{color}

\begin{document}

\title{Stability limits for modes held in alternating trapping-expulsive
potentials}
\author{Zhihuan Luo$^{1}$, Yan Liu$^{1}$, Yongyao Li$^{2}$, Josep Batle$^{3}$%
, and Boris A. Malomed$^{4,5}$}
\affiliation{$^{1}$ Department of Applied Physics, South China Agricultural University,
Guangzhou 510642, China \\
$^{2}$ School of Physics and Optoelectronic Engineering, Foshan University,
Foshan $528000$, China \\
$^{3}$ CRISP Centre de Recerca Independent de sa Pobla, C. Alb\'eniz 12,
07420 sa Pobla, Balearic Islands, Spain \\
$^{4}$ Department of Physical Electronics, School of Electrical Engineering,
Faculty of Engineering, and Center for Light-Matter Interaction, Tel Aviv
University, Tel Aviv 69978, Israel\\
$^{5}$Instituto de Alta Investigaci\'{o}n, Universidad de Tarapac\'{a},
Casilla 7D, Arica, Chile}

\begin{abstract}
We elaborate a scheme of trapping-expulsion management (TEM), in the form of
the quadratic potential periodically switching between confinement and
expulsion, as a means of stabilization of two-dimensional dynamical states
against the backdrop of the critical collapse driven by the cubic
self-attraction with strength $g$. The TEM scheme may be implemented, as
spatially or temporally periodic modulations, in optics or BEC,
respectively. The consideration is carried out by dint of numerical
simulations and variational approximation (VA). In terms of the VA, the
dynamics amounts to a nonlinear Ermakov equation, which, in turn, is
tantamount to a linear Mathieu equation. Stability boundaries are found as
functions of $g$ and parameters of the periodic modulation of the trapping
potential. Below the usual collapse threshold, which is known, in the
numerical form, as $g<g_{\mathrm{c}}^{(\mathrm{num})}\approx 5.85$ (in the
standard notation), the stability is limited by the onset of the parametric
resonance. This stability limit, including the setup with the self-repulsive
sign of the cubic term ($g<0$), is accurately predicted by the VA. At $g>g_{%
\mathrm{c}}^{(\mathrm{num})}$, the collapse threshold is found with the help
of full numerical simulations. The relative increase of $g_{\mathrm{c}}$
above $g_{\mathrm{c}}^{(\mathrm{num})}$ is $\approx 1.5\%$, which is a
meaningful result, even if its size is small, because the collapse threshold
is a universal constant which is difficult to change.
%This result is valid because oscillatory bound states (breathers)
%remain immune to the collapse at values of $g$ which are somewhat larger
%than $g_{\mathrm{c}}^{(\mathrm{num})}$, which pertains to stationary TSs.

%\vspace{6pt} \textbf{\emph{Keywords}}: Bose-Einstein condensates;
%multidimensional solitons; trapping potential; management techniques;
%variational approximation; Ermakov equation; Mathieu equation
\end{abstract}

\maketitle

%\email{malomed@post.tau.ac.il}

\section{ Introduction}

It is well known that two- and three-dimensional (2D and 3D)
multidimensional solitons, maintained by the ubiquitous cubic
self-attraction, are subject to severe instabilities, due to the fact that
the same nonlinearity drives the critical and supercritical collapse, in the
2D and 3D cases, respectively \cite{Berge,Sulem,Fibich}. The search for
physically relevant settings which make it possible to stabilize
self-trapped multidimensional states is a relevant problem, especially in
the context of nonlinear optics and matter-wave patterns in Bose-Einstein
condensates (BECs). Methods elaborated for this purpose include the use of
quadratic interactions \cite{quadratic}, higher-order defocusing
nonlinearity, which may be represented by quintic terms, that occur in
optics \cite{Quiroga,Dimitrevski,Pego,Angers,9authors,Michinel,Recife}, and
quartic ones, that account for the stabilization of\textit{\ quantum droplets%
} by quantum fluctuations in binary BEC \cite{Petrov1,Petrov2,swirling}),
spin-orbit coupling (SOC) acting on binary condensates \cite%
{Sakaguchi-2014-PRE1,Salasnich-2014-PRA,Sherman}, etc., see reviews \cite%
{Kivshar-2000-PR,Malomed-2005-JPB,Smyth,Malomed-2016-EPJ,NatRevPhys,Luo-2021-FoP}%
. Experimentally, soon after the theoretical prediction of quantum droplets,
they have been created in BEC with local \cite%
{Cabrera-Sci-2018,Cheiney-PRL-2018,Semeghini-PRL-2018,Ferioli-PRL-2019,
Errico-PRR-2019} and nonlocal \cite%
{FB-PRL-2016,Schmitt-Nature-2016,FB-2018-PRL,Chomaz-PRX-2016} interactions.

Straightforward means for the stabilization of 2D self-attractive fields
with zero vorticity (i.e., fundamental states, FSs) and vorticity $S=1$
against the critical collapse is provided by the harmonic-oscillator (HO)
trapping potential \cite{Dalfovo-1996-PRA, Adhikari-2001-PRE,Alexander,
Saito-2002-PRL, Saito-2004-PRA, Mihalache-2006-PRA,Carr, Boris-2007-PLA}. In
particular, it has been found that the 2D collapse instability of all FS
modes is completely removed by the HO potential, while the vortex modes
remain unstable against spontaneous splitting, only the ones with $S=1$ and
the norm falling below a certain threshold value being stabilized by the
trapping potential. In addition to that, in an interval of values of the
norm above the threshold there are stable dynamical states, in the form of
vortices with $S=1$ which periodically split in two fragments and recombine
back, keeping the angular momentum.

On the other hand, expulsive quadratic (\textit{anti-HO}) potentials also
appear in a variety of physically relevant setups \cite{Carr2}-\cite{KartKon}%
. In guided-wave optics, expulsive potentials occur in anti-waveguiding
systems, which are used to design various data-processing photonic schemes
\cite{anti-WG1}-\cite{anti-WG4}. The latter application makes it relevant to
consider the propagation of optical solitons through a waveguide built of
alternating trapping and expulsive segments \cite{Kaplan}. A similar setting
is possible in BEC, with the external potential periodically \cite%
{Galimzyanov} or temporarily \cite{Muga,cooling} switching between the HO
and anti-HO forms. It may be realized experimentally, using the usual
optical trapping setup for BEC \cite{Grimm,Bagnato}, pumped by modulated
light, which periodically switches between red- and blue-detuned
frequencies. Search for stable solitons existing under the joint action of
the cubic self-attractive nonlinearity and periodically alternating
trapping-expulsive potential places such settings in the class of systems
which maintain solitons by means of various \textit{management} techniques
\cite{book}, a commonly known example being \textit{dispersion management}
of temporal solitons in optical fibers \cite{Turitsyn}.

However, previous works considered the interplay of the self-focusing and
periodically flipping HO -- anti-HO potential only in 1D geometry. The
objective of the present work is to develop this analysis for 2D solitons.
This is a challenging problem because of the possibility of the critical
collapse (blowup) in such a case. This circumstance links the present
problem to the setting based on the \textit{nonlinearity management} for 2D
solitons in free space (in the absence of the trapping potential), which was
originally introduced in optics, considering the propagation of (2+1)D
spatial solitons in bulk waveguides built as alternation of layers with
self-focusing and defocusing Kerr nonlinearity \cite{Isaac}. Later it was
extended for BEC\ in the quasi-2D geometry, with the sign of the contact
nonlinearity periodically switching between attraction and repulsion under
the action of the Feshbach resonance controlled by a periodically varying
magnetic field \cite{Fatkh,Ueda,Itin}.

In the case of the critical collapse, modelled by the 2D nonlinear Schr\"{o}%
dinger equation [alias the Gross-Pitaevskii equation (GPE), in terms of the
mean-field description of BEC \cite{Pit}], a crucially important role is
played by Townes solitons (TSs) \cite{Townes}, which realize the separatrix
between decaying and collapsing solutions of the GPE in 2D. As any
separatrix solution \cite{sep1,sep2}, the TSs are unstable against small
perturbations. In the free space, the TS family is degenerate in the sense
that all solitons belonging to it have a single value of the norm, $N_{%
\mathrm{TS}}$. This value represents a threshold necessary for the onset of
the collapse, as, at the late stage of the blowup, the collapsing mode
becomes very narrow, hence the collapse ends up as in the nearly-free space,
even if an external potential is present. The system with the fully unstable
TS\ family has no ground state (it is replaced by the collapsing one). In
fact, the stabilization of the TSs by the trapping potential \cite%
{Dalfovo-1996-PRA,Alexander,Mihalache-2006-PRA} or SOC \cite%
{Sakaguchi-2014-PRE1,Sherman} is underlain by the fact that the potential or
SOC lifts the degeneracy, making it possible to create FS modes with $N<N_{%
\mathrm{TS}}$, which are stable because the collapse does not take place in
this case. However, such results did not demonstrate a possibility to
stabilize states with $N>N_{\mathrm{TS}}$. An essential result of the
present work is that the application of the \textquotedblleft
trapping-expulsive management" (TEM) makes it possible to construct stable
FS modes, in an oscillatory form, whose norm exceeds $N_{\mathrm{TS}}$ by a
small but meaningful margin, $\approx 1.5\%$, while usually $N_{\mathrm{TS}}$
is a universal constant, which cannot be changed. Another major objective of
the work is to identify robustness boundaries of the FS modes under the
action of the TEM at $N<N_{\mathrm{TS}}$ against the action of a different
potentially destabilizing factor -- not the collapse, but the parametric
resonance (PR), which may occur when an internal mode of the FS trapped in
the potential resonates with the TEM frequency.

The rest of this paper is structured as follows. The model is introduced in
Section II. In that section, physical parameters for the realization of the
model in BEC and optics are evaluated too. The variational approximation
(VA), which provides a relevant method to predict the stability of the FSs
in the present model, is also elaborated in Section II. The VA replaces the
GPE by a second-order nonlinear ODE of the \textit{Ermakov's type} (see
details below). The latter equation is simulated numerically, but the onset
of the instability, caused by the PR, is correctly predicted in an
analytical form, as the Ermakov equation is tantamount to the linear Mathieu
equation. Results of the systematic numerical investigation, which
demonstrate stability boundaries of the FSs against the critical collapse
and PR alike, are summarized in Section III. In particular, the VA predicts
the PR-instability boundary very accurately, including the system with the
self-repulsive nonlinearity. The paper is concluded by Section IV.

\section{The model and variational approximation (VA)}

We adopt the single-component GPE, written in the scaled 2D form for the
mean-field BEC wave function, $\psi $:

\begin{equation}
i\frac{\partial \psi }{\partial t}=-\frac{1}{2}\nabla ^{2}\psi -g|\psi
|^{2}\psi +\frac{1}{2}\kappa (t)r^{2}\psi ,  \label{eq-model2}
\end{equation}%
where $g>0$ is the constant coefficient of the cubic self-attraction and $r$
is the radial coordinate. TEM is introduced by making the strength of the
quadratic potential a function of time, which includes dc (constant) and ac
(variable) components, periodically flipping between positive and negative
values:

\begin{equation}
\kappa (t)=\kappa _{\mathrm{dc}}+\kappa _{\mathrm{ac}}\cos \left( \omega
t\right) .  \label{eq-kappa}
\end{equation}%
The case of basic interest is
\begin{equation}
\kappa _{\mathrm{ac}}>\kappa _{\mathrm{dc}}>0,  \label{>>0}
\end{equation}%
as this condition maintains the sign-changing structure of function (\ref%
{eq-kappa}). The case of
\begin{equation}
0<\kappa _{\mathrm{ac}}<\kappa _{\mathrm{dc}}  \label{0<<}
\end{equation}
is briefly considered below too.

In terms of optics, temporal variable $t$ in Eqs. (\ref{eq-model2}) and (\ref%
{eq-kappa}) is replaced by the propagation distance, $z$, $g>0$ is the
scaled Kerr coefficient, and coefficient $\kappa (z)$ represents the
guiding-antiguiding structure in the bulk material. In that case, a more
realistic form of the periodic modulation is piecewise-constant, with
spatial period $2\pi /\omega $, while the $\cos $ term in Eq. (\ref{eq-kappa}%
) represents its first harmonic.

In spite of the presence of the time dependence in Eq. (\ref{eq-model2}), it
conserves two dynamical invariants, \textit{viz}., the norm, proportional to
the number of atoms in BEC (or the integral power, in terms of optics),%
\begin{equation}
N=\int \int dxdy\left\vert \psi (x,y)\right\vert ^{2},  \label{norm}
\end{equation}%
and the angular momentum,%
\begin{equation}
M=i\int \int dxdy\psi ^{\ast }\left( y\frac{\partial }{\partial x}-x\frac{%
\partial }{\partial y}\right) \psi ,  \label{M}
\end{equation}%
where $\ast $ stands for the complex conjugate. In this work, only states
with $M=0$ are considered. These invariants correspond, respectively, to the
invariance of Eq. (\ref{eq-model2}) with respect to a phase shift of the
wave function, and rotation of coordinate system $\left( x,y\right) $.

As concerns the realization of the present model in BEC, an estimate for the
atomic condensate of $^{7}$Li atoms, with the scattering length $\simeq -0.1$
nm accounting for the attractive interactions, and the confinement length in
the transverse direction $\simeq 1$ $\mathrm{\mu }$m (which implies the
trapping frequency $\simeq 10$ kHz), shows that the length and time units in
the scaled variables used in Eq. (\ref{eq-model2}) correspond, respectively,
to $\simeq 10$ $\mathrm{\mu }$m (which implies the trapping frequency $%
\simeq 10$ kHz) and $100$ ms in physical units, cf. Ref. \cite{Viskol}. In
this case, the critical number of atoms leading to the collapse is estimated
as $6\times 10^{3}$.

In terms of optical waveguides, the use of the carrier wavelength $600$ nm
in silica leads to a conclusion that the units of the propagation distance
and transverse coordinates in scaled equation (\ref{eq-model2}) typically
correspond to $\simeq 1$ mm and $15$ $\mathrm{\mu }$m, respectively, in
physical units (cf. Ref. \cite{spatial soliton}). The corresponding total
power of the optical beam may be estimated as $\simeq 3$ MW.

The main issue addressed in this work is to identify conditions under which
the TEM scheme based on Eqs. (\ref{eq-model2}) and (\ref{eq-kappa}) is able
to hold $\psi $ in a robust dynamical state, preventing both the collapse
and decay. As mentioned above, this issue is somewhat similar to the problem
of identifying conditions for holding a stable 2D soliton by the GPE with
periodically sign-flipping self-interaction coefficient, representing the
nonlinearity management:%
\begin{equation}
i\frac{\partial \psi }{\partial t}=-\frac{1}{2}\nabla ^{2}\psi -g(t)|\psi
|^{2}\psi ,  \label{g}
\end{equation}%
with $g(t)=g_{0}+g_{1}\cos \left( \omega t\right) $, $g_{1}>g_{0}>0$, cf.
Eqs. (\ref{eq-kappa}) and (\ref{>>0}) \cite{Fatkh,Ueda,Itin}. The stability
area for 2D quasi-Townes solitons with zero vorticity [i.e., FSs, for which
Eq. (\ref{M}) yields $M=0$] was identified in the latter model, while all
states with nonzero vorticity are unstable (in a two-component system with
cross-attraction, a vortex soliton in one component may be stabilized by the
FS in the other one \cite{cross-vortex}). Here, we do not consider vortex
states governed by Eq. (\ref{eq-model2}), as they should be a subject of a
separate work.

Equation (\ref{eq-model2}) can be derived from the Lagrangian,%
\begin{equation}
L=\frac{1}{2}\int \int dxdy\left[ i\left( \psi ^{\ast }\frac{\partial \psi }{%
\partial t}+\mathrm{c.c.}\right) -\left\vert \nabla \psi \right\vert
^{2}+g|\psi |^{4}-\kappa (t)r^{2}|\psi |^{2}\right] ,  \label{L}
\end{equation}%
where c.c. stands for the complex-conjugate expression. Following Ref. \cite%
{Anderson}, the VA can be based on the usual Gaussian ansatz,%
\begin{equation}
\psi _{\mathrm{ans}}\left( r,t\right) =A(t)\exp \left[ -\frac{r^{2}}{%
2W^{2}(t)}+i\phi (t)+ib(t)r^{2}\right] ,  \label{ans}
\end{equation}%
where real variational parameters are amplitude $A(t)$, width $W(t)$, radial
chirp $b(t)$, and phase $\phi (t)$. The conserved norm (\ref{norm}) of the
ansatz is%
\begin{equation}
N=\pi A^{2}W^{2}.  \label{N}
\end{equation}

The substitution of ansatz (\ref{ans}) in Lagrangian (\ref{L}) and
integration yields the VA Lagrangian:%
\begin{equation}
L_{\mathrm{VA}}=-NW^{2}\frac{db}{dt}-N\left( 2W^{2}b^{2}+\frac{1}{2W^{2}}%
\right) +\frac{gN^{2}}{4\pi W^{2}}-\frac{1}{2}\kappa (t)NW^{2}.  \label{LVA}
\end{equation}%
To derive this expression, Eq. (\ref{N}) was used to eliminate $A^{2}$ in
favor of $W$. The first variational (Euler-Lagrange) equation, $\delta L_{%
\mathrm{VA}}/\delta b=0$, applied to Lagrangian (\ref{LVA}), yields a
relation which expresses the chirp in terms of $W(t)$:%
\begin{equation}
b=\frac{1}{2W}\frac{dW}{dt}.  \label{b}
\end{equation}%
The second Euler-Lagrange equation, $\partial L_{\mathrm{VA}}/\partial W=0$,
produces the final dynamical equation, in which Eq. (\ref{b}) was used to
eliminate $b$:%
\begin{equation}
\frac{d^{2}W}{dt^{2}}=\left( 1-\frac{gN}{2\pi }\right) \frac{1}{W^{3}}-\left[
\kappa _{\mathrm{dc}}+\kappa _{\mathrm{ac}}\cos \left( \omega t\right) %
\right] W.  \label{eq-W}
\end{equation}%
The TEM term $\sim \kappa _{\mathrm{ac}}$ plays the role of the \textit{%
parametric drive} in Eq. (\ref{eq-W}).

The coefficient in front of term $1/W^{3}$ in Eq. (\ref{eq-W}) vanishes at
the critical point,%
\begin{equation}
\left( gN\right) _{\mathrm{c}}^{\mathrm{(VA)}}=2\pi .  \label{crit}
\end{equation}%
This value is well known as the VA prediction for the TS norm \cite{Anderson}%
, whose numerically found value is $\simeq 7\%$ smaller (\ref{crit}):
\begin{equation}
\left( gN\right) _{\mathrm{c}}^{\mathrm{(num)}}\approx 5.85.  \label{Townes}
\end{equation}%
From now on, we fix, by means of rescaling,%
\begin{equation}
N\equiv 1,\kappa _{\mathrm{dc}}\equiv 1  \label{1}
\end{equation}%
(unless $\kappa _{\mathrm{dc}}=0$ is fixed in some cases, see below), hence
Eqs. (\ref{crit}) and (\ref{Townes}) determine the critical values of the
self-attraction strength, above which the critical collapse is initiated in
the system,%
\begin{equation}
g_{\mathrm{c}}^{\mathrm{(VA)}}=2\pi ,g_{\mathrm{c}}^{\mathrm{(num)}}\approx
5.85.~  \label{gg}
\end{equation}

In the case of $g<g_{\mathrm{c}}^{\mathrm{(VA)}}$, the constant solution
(alias the fixed point, FP) of the stationary version of Eq. (\ref{eq-W}),
with $\kappa _{\mathrm{ac}}=0$, is
\begin{equation}
W_{\mathrm{FP}}=\left( {1-}\frac{{g}}{2\pi }\right) ^{1/4},
\label{eq-W0stab}
\end{equation}%
where normalization (\ref{1}) is taken into regard. This solution is
obviously stable, as it realizes a minimum of the respective Hamiltonian, in
the case of $\kappa _{\mathrm{ac}}=0$. The frequency of small oscillations
around the FP does not depend on $g$,%
\begin{equation}
\Omega _{\mathrm{FP}}=2,  \label{2}
\end{equation}%
as long as it falls below the critical value, $g<2\pi $, see Eq. (\ref{gg}).

This analysis can be readily extended to the case of $\kappa _{\mathrm{ac}%
}\neq 0$, provided that the modulation frequency $\omega $ is large, by
means of the averaging method, cf. Ref. \cite{Galimzyanov}. In this case, an
approximate solution to Eq. (\ref{eq-W}) is looked for as
\begin{equation}
W(t)=W^{(0)}(t)+W^{(1)}\cos \left( \omega t\right) ,  \label{WWW}
\end{equation}%
where $W^{(0)}(t)$ is a slowly varying term, and the harmonic balance yields%
\begin{equation}
W^{(1)}=\left( \kappa _{\mathrm{ac}}/\omega ^{2}\right) W^{(0)}.  \label{WW}
\end{equation}%
Then, the substitution of expressions (\ref{WWW}) and (\ref{WW}) in Eq. (\ref%
{eq-W}) leads to an effective equation for the slow evolution of $W^{(0)}(t)$%
:%
\begin{equation}
\frac{d^{2}W^{(0)}}{dt^{2}}=\left( 1-\frac{gN}{2\pi }\right) \frac{1}{\left(
W^{(0)}\right) ^{3}}-\left( 1+\frac{\kappa _{\mathrm{ac}}^{2}}{2\omega ^{2}}%
\right) W^{(0)}.  \label{slow}
\end{equation}%
The respective FP value changes from the one given by Eq. (\ref{eq-W0stab})
to%
\begin{equation}
W_{\mathrm{FP}}\approx \left( {1-}\frac{{g}}{2\pi }\right) ^{1/4}\left( 1-%
\frac{\kappa _{\mathrm{ac}}^{2}}{8\omega ^{2}}\right) ,  \label{FPaveraged}
\end{equation}%
and frequency (\ref{2}) of small oscillations around the FP is replaced by%
\begin{equation}
\Omega _{\mathrm{FP}}\approx 2\left( 1+\frac{\kappa _{\mathrm{ac}}^{2}}{%
4\omega ^{2}}\right) ,  \label{2shifted}
\end{equation}%
where the smallness of $\omega ^{-2}$ is taken into regard. On the other
hand, the correction induced by rapid oscillations does not affect the
VA-predicted collapse threshold given by Eq. (\ref{gg})

The nonexistence of solution (\ref{eq-W0stab}) at $g>2\pi$
in the case of $\kappa _{\mathrm{ac}}=0$ signals the transition to the
collapse. In this case, the onset of the collapse is described by a simple
solution of Eq. (\ref{eq-W}) with $\kappa _{\mathrm{dc,ac}}=0$ (recall $N=1$
is fixed):%
\begin{equation}
W(z)=W_{0}\sqrt{1-t/t_{\text{coll}}},  \label{W(z)}
\end{equation}%
where $W_{0}$ is the initial width, and the VA-predicted collapse time is%
\begin{equation}
t_{\mathrm{coll}}=\frac{2W_{0}^{2}}{\sqrt{g/(2\pi )-1}}  \label{tcoll}
\end{equation}

If the action of the ac component in Eq. (\ref{eq-W}) gives rise to
instability through excitation of the PR, development of the instability
implies that the amplitude of oscillations of $W(t)$ grows, hence the term $%
\sim W^{-3}$ in Eq. (\ref{eq-W}) becomes negligible. The corresponding
linear equation is the classical Mathieu equation \cite{Mathieu}:%
\begin{equation}
\frac{d^{2}W}{dt^{2}}=-\left[ \kappa _{\mathrm{dc}}+\kappa _{\mathrm{ac}%
}\cos \left( \omega t\right) \right] W.  \label{Mathieu}
\end{equation}%
The commonly known instability chart of Eq. (\ref{Mathieu}) in the plane of $%
\left( \kappa _{\mathrm{ac}},\omega \right) $ (see, e.g., Ref. \cite%
{MW-2013-AIM}) is determined by the fundamental and higher-order PRs \cite%
{LL}. In the limit of $\kappa _{\mathrm{ac}}\rightarrow 0$, the PRs of
orders $m=0,1,2,...$ take place at values of the driving frequency%
\begin{equation}
\omega _{\mathrm{PR}}^{(n)}=2/\left( 1+m\right) ,  \label{param-res}
\end{equation}%
the fundamental (strongest) PR corresponding to $m=0$, i.e., $\omega _{%
\mathrm{PR}}^{(0)}=2$.

The relation between Eqs. (\ref{eq-W}) and (\ref{Mathieu}) is not
surprising, as Eq. (\ref{eq-W}) belongs to the class of the Ermakov's
equations\textit{\ }\cite{Ermakov,commentary,Lewis1,Lewis,Lohe}. In this
context, it is well known that a general solution of Eq. (\ref{eq-W}), $W(t)$%
, may be exactly expressed in terms of two independent solutions, $w_{1}(t)$
and $w_{2}(t)$, of the Mathieu equation (\ref{Mathieu}), and their constant
Wronskian, as follows:%
\begin{equation}
\frac{W^{2}(t)}{\sqrt{1-g/(2\pi )}}=w_{1}^{2}(t)+\left( \frac{w_{2}(t)}{%
\mathrm{Wronskian}\left\{ w_{1}(t),w_{2}(t)\right\} }\right) ^{2}.
\label{Wronskian}
\end{equation}%
Thus, Eq. (\ref{Wronskian}) corroborates that the onset of the instability
in solutions of the Mathieu equation implies, in the exact form, that the
solutions of the Ermakov equation (\ref{eq-W}) also becomes unstable.

\section{Numerical results}

\subsection{The formulation of the problems}

First, stationary solutions of Eq. (\ref{eq-model2}), without the ac drive ($%
\kappa _{\mathrm{ac}}=0$) and with $g<g_{\mathrm{c}}^{\mathrm{(num)}}\approx
5.85$ [see Eq. (\ref{gg})], were produced numerically by means of the
well-known imaginary-time integration method \cite{Im-time1,Im-time2,Yang}.
Then, using the stationary solutions as inputs, we performed simulations of
GPE (\ref{eq-model2}) in the full form, including the ac drive, by means of
the standard split-step fast-Fourier-transform algorithm \cite{Yang}. The
drive's parameters, $\kappa _{\mathrm{ac}}$ and $\omega $, were varied with
the aim to identify stable and unstable dynamical states in the course of
the long-time evolution. One of the main objectives of this work being to
explore the possibility of finding stable ac-driven FSs at $g>g_{\mathrm{crit%
}}^{\mathrm{(num)}}$, when no stationary solution exists in the absence of
the ac drive, the input in this case was taken as the normalized Gaussian,%
\begin{equation}
\psi \left( r,t=0\right) =\sqrt{1/\pi }\exp \left( -r^{2}/2\right) .
\label{input}
\end{equation}%
The simulations were performed in the domain of size $12\times 12$, with the
spatial mesh size $\Delta x=\Delta y=0.03$ (i.e., the integration domain was
covered by the mesh composed of $400\times 400$ points) and the time step $%
\Delta t=0.0001$. A boundary absorber was inserted in the simulations, to
prevent irrelevant perturbation of the dynamical state by waves reflected
from the domain's boundary.

The dynamics of the FS mode under the action of TEM may be adequately
characterized by time dependences of its height and width, i.e., the (peak)
density at the center, $\left\vert \psi \left( r=0,t\right) \right\vert ^{2}$%
, and the monopole moment, which determines the average radial size of the
mode,%
\begin{equation}
\left\langle r\right\rangle (t)=N^{-1}\int \int |\psi \left( x,y\right) |^{2}%
\sqrt{x^{2}+y^{2}}dxdy  \label{r}
\end{equation}%
[in fact, we set $N\equiv 1$, see Eq. (\ref{1})]. Also essential are spectra
of the Fourier transform of the central density and radial size, computed as%
\begin{equation}
\hat{n}(\Omega )=\left\vert \int_{0}^{T}e^{-i\Omega t}\left\vert \psi \left(
r=0,t\right) \right\vert ^{2}dt\right\vert ,\hat{r}\left( \Omega \right)
=\left\vert \int_{0}^{T}e^{-i\Omega t}\left\langle r\right\rangle
(t)dt\right\vert ,  \label{Fourier}
\end{equation}%
for a long simulation interval $T$. Note that, in terms of the VA ansatz,
represented above by Eqs. (\ref{ans}), (\ref{N}), and (\ref{1}), these
characteristics are given by%
\begin{equation}
\left\vert \psi \left( r=0,t\right) \right\vert ^{2}=\left( \pi W^{2}\right)
^{-1},\left\langle r\right\rangle =\left( \sqrt{\pi }/2\right) W.
\label{VA-comparison}
\end{equation}

The boundary of the PR-induced instability, produced by systematic
simulations of GPE (\ref{eq-model2}), is then compared to its counterpart
predicted by the VA, based on simulations of the Ermakov equation (\ref{eq-W}%
). Those long-time simulations start with the input taken as per Eq. (\ref%
{eq-W0stab}).

\subsection{The stability boundary of the FS (fundamental state)\ against
the PR\ (parametric resonance)\ under the action of the TEM\
(trapping-expulsion management)}

\begin{figure}[tbp]
\centering{\includegraphics[scale=0.28]{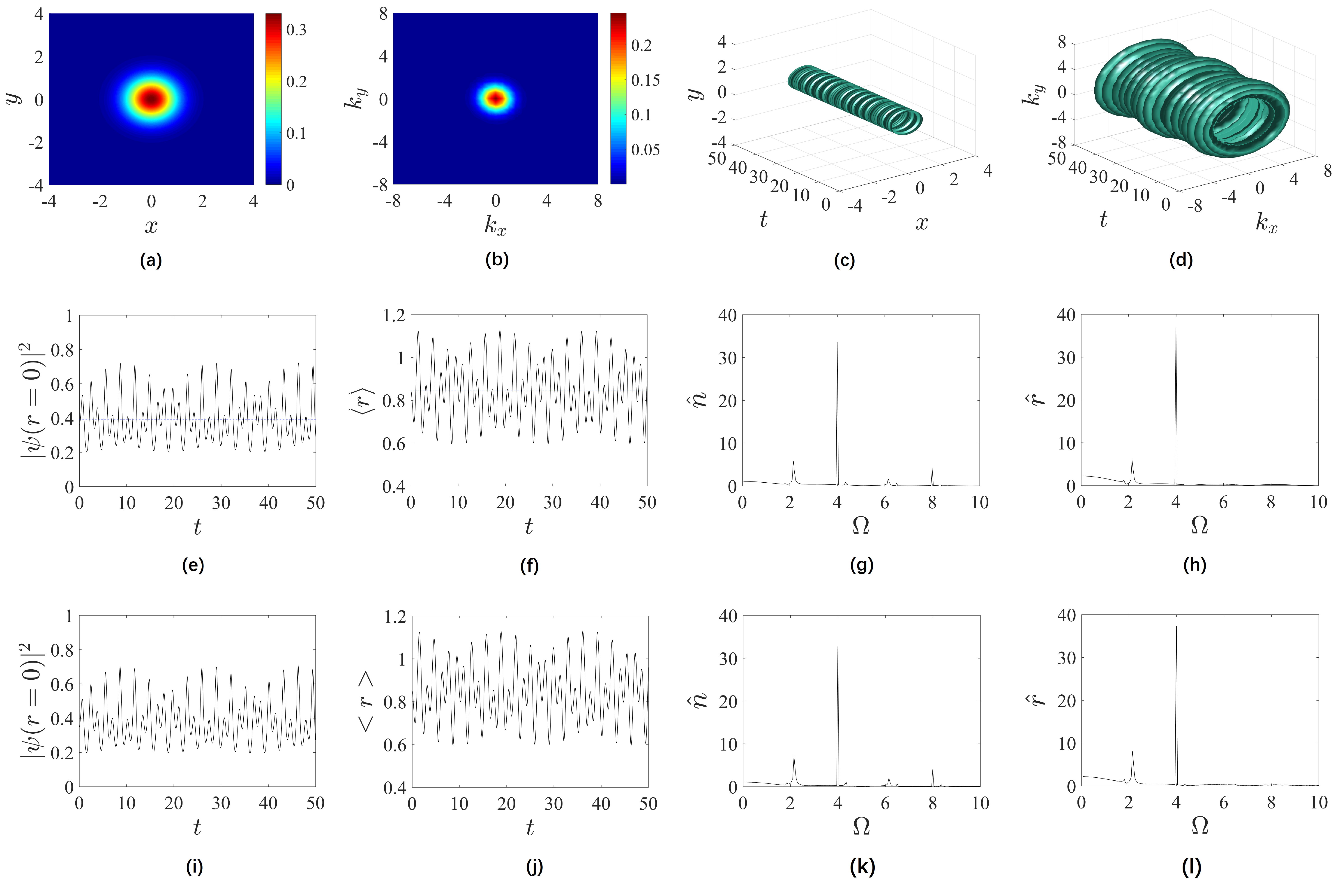}}
\caption{(Color online) An example of stable dynamics of the FS (fundamental
state) governed by Eqs. (\protect\ref{eq-model2}) and (\protect\ref{1}),
with $g=1$, $\protect\kappa _{\mathrm{ac}}=2$ and $\protect\omega =4$. (a)
The input density pattern (\protect\ref{n}), taken as per the stationary
solution of Eq. (\protect\ref{eq-model2} with $\protect\kappa \equiv 1$. (b)
The spatial Fourier transform of (a), defined according to Eq. (\protect\ref%
{nFourier}). (c,d) The evolution of the density profile and its spatial
Fourier transform. (e,f) The corresponding evolution of the peak density, $|%
\protect\psi (r=0,t)|^{2}\equiv n(t)$, and monopole moment (radial size),
defined as per Eq. (\protect\ref{r}). Dashed horizontal lines in panels (e)
and (f) show average values of the corresponding variables. (g,h) The
temporal Fourier transform of the peak density, $\hat{n}(\Omega )$, and
radial size, $\hat{r}(\Omega )$, calculated as per Eq. (\protect\ref{Fourier}%
) with $T=500$. Panels (i-l) demonstrate the results corresponding to those
in (e-h), as produced by the VA.}
\label{fig01}
\end{figure}

First, Fig. \ref{fig01} represents a typical example of fully robust
evolution of the FS under the action of TEM, with $\omega =4$ and $\kappa _{%
\mathrm{ac}}=2$ exceeding $1$, hence the sign of the quadratic potential in
Eq. (\ref{eq-model2}) indeed periodically flips, according to Eq. (\ref%
{eq-kappa}). This figure is produced for $g=1$, which is far from the
critical value (\ref{gg}). Panels (a) and (b) display the pattern of the
local density in the input [stationary solution of Eq. (\ref{eq-model2})
with $\kappa _{\mathrm{ac}}=0$],%
\begin{equation}
n\left( x,y\right) =\left\vert \psi \left( x,y\right) \right\vert ^{2},
\label{n}
\end{equation}%
and its Fourier transform,%
\begin{equation}
\hat{n}\left( k_{x},k_{y}\right) =\int \int \exp \left( -i\left(
k_{x}x+k_{y}y\right) \right) n\left( x,y\right) dxdy\equiv 2\pi
\int_{0}^{\infty }n(r)J_{0}\left( kr\right) rdr,  \label{nFourier}
\end{equation}%
where it is taken into regard that $n$ depends only on the radial
coordinate, $k=\sqrt{k_{x}^{2}+k_{y}^{2}}$, and $J_{0}$ is the Bessel
function. Further, the periodic evolution of the density and its Fourier
transform in the numerically generated solution with $\kappa _{\mathrm{ac}%
}=2 $ is displayed in panels (c) and (d), respectively.

Naturally, the peak density, $n\left( r=0\right) $, and radial size (\ref{r}%
) of the solution with $\kappa _{\mathrm{ac}}=2$ feature anti-phase
oscillations in panels \ref{fig01}(e) and (f). The spectrum of the
oscillations, displayed in panels \ref{fig01}(g) and (h), features the main
peak at the driving frequency, $\omega =4$, a weak subharmonic peak at $%
\omega =2$, a weak one at the double frequency, $\omega =8$, and an
additional very weak but visible peak at the combinational sesquilateral
harmonic, $\omega =6$. Furthermore, comparison of panels (i)-(l) and (e)-(h)
demonstrates close agreement between the simulations of the full GPE and
their VA-produced counterparts. It has been also checked that, in the case
of the stability, full simulations of the underlying GPE always conserve the
total norm (\ref{N}) of the wave function.

\begin{figure}[tbp]
\centering{\includegraphics[scale=0.3]{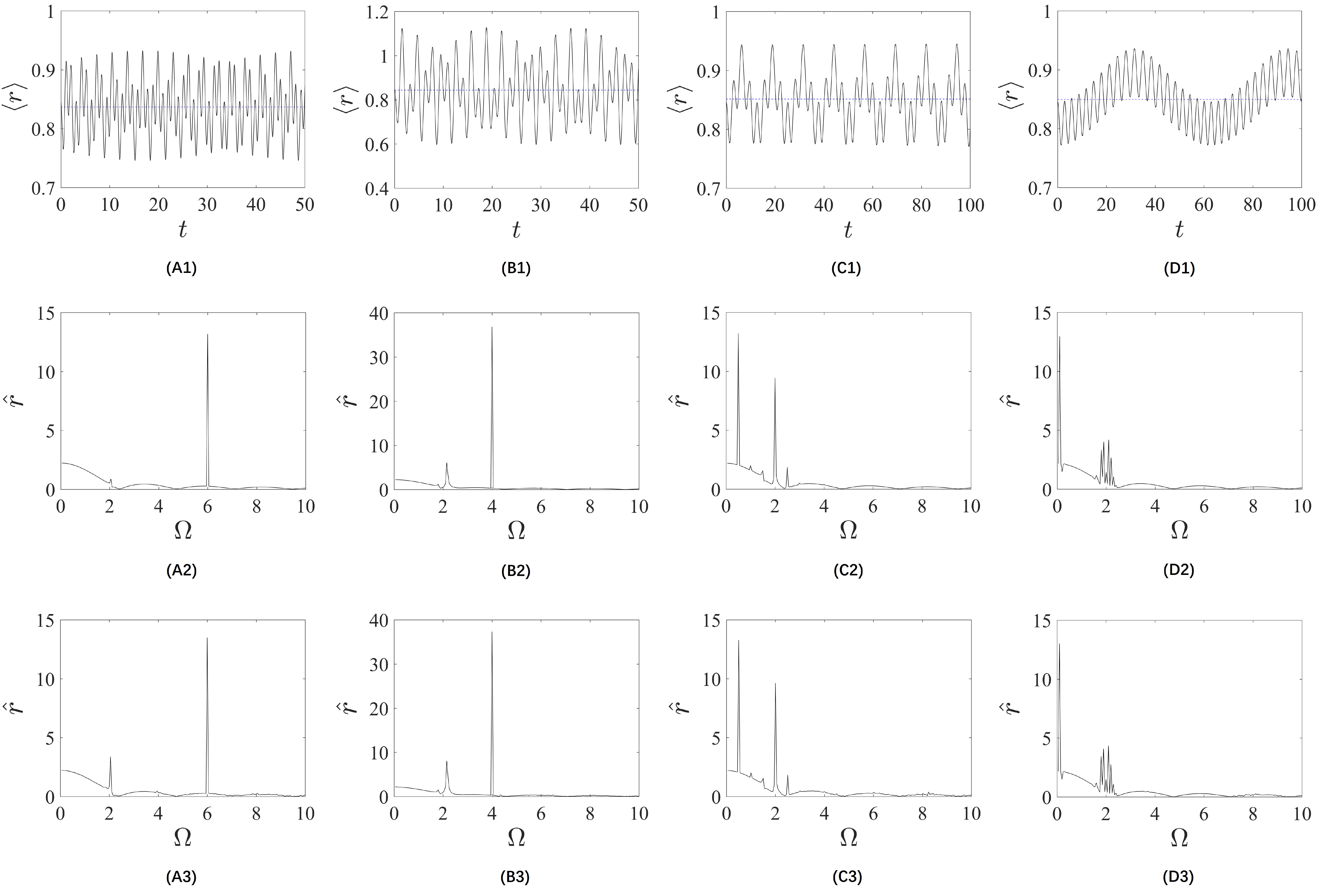}}
\caption{Panels (A1)-(D1): the radial size [monopole moment, see Eq. (%
\protect\ref{r})] vs. time, produced by the simulations of Eq. (\protect\ref%
{eq-model2}). (A2)-(D2): the respective spectra. (A3)-(D3): the spectra as
produced by the VA. The parameters are $(\protect\kappa _{\mathrm{ac}},%
\protect\omega )=(2,6)$ in (A1)-(A3); $(2,4)$ in (B1)-(B3); $(0.2,0.5)$ in
(C1-C3); and $(0.2,0.1)$ in (D1)-(D3). In all plots, $(g,\protect\kappa _{%
\mathrm{dc}})=(1,1)$. At $\protect\omega >2$, the main peak in the spectra,
produced by the GPE and VA alike, is at $\Omega =\protect\omega $. At $%
\protect\omega <2$, there is an additional major spectral peak close to $%
\Omega =2$ [it is split in subpeaks in panels (D2) and (D3)].}
\label{fig02}
\end{figure}

The structure of the oscillatory states shown in Fig. \ref{fig01} is typical
for relatively large values of the driving frequency $\omega $. At small
values of $\omega $, it may be essentially different: in addition to the
main low-frequency peak at $\Omega =\omega $, the spectrum features a
somewhat weaker but conspicuous one at a much larger frequency, $\Omega
\simeq 2$, which is easily predicted by Eq. (\ref{2}), in the case when $%
\omega $ is small. Examples of this are displayed in Fig. \ref{fig02}, in
which the top and middle rows show, respectively, the time dependence of the
FS's radial size, $\left\langle r\right\rangle (t)$, and its spectral
counterpart, $\hat{r}(\Omega )$, as obtained from the simulations of GPE (%
\ref{eq-model2}). In addition, the bottom row shows $\hat{r}(\Omega )$ as
produced by the VA. Typical examples of the low-frequency cases, with $%
\left( \omega ,\kappa _{\mathrm{ac}}\right) =\left( 0.5,0.2\right) $ and $%
\left( \omega ,\kappa _{\mathrm{ac}}\right) =\left( 0.1,0.2\right) $, are
presented, respectively, in columns (C1)-(C3) and (D1)-(D3). Note that the
peak at $\Omega \simeq 2$ in panels (D2) and (D3) is split into
subcomponents by combinations of the main one with those corresponding to
the small driving frequency $\omega $ [a weak combinational peak at $\Omega
=2+\omega $ is observed as well in panels (C2) and (C3)]. The oscillatory
states displayed in columns (C1)-(C3) and (D1)-(D3) are akin to breathers
which are represented, in the framework of the VA, by oscillatory solutions
of Eq. (\ref{eq-W}) in the absence of the ac drive, i.e., with $\kappa _{%
\mathrm{ac}}=0$.

For the comparison's sake, columns (A1)-(A3) and (B1)-(B3) in Fig. \ref%
{fig02} represent typical high-frequency cases, with $\left( \omega ,\kappa
_{\mathrm{ac}}\right) =\left( 2,6\right) $ and $\left( \omega ,\kappa _{%
\mathrm{ac}}\right) =\left( 2,4\right) $, respectively. Note that the
corresponding spectra also include peaks at $\Omega \approx 2$, although
small-amplitude ones. Furthermore, a small but visible shift of these peaks
to values of $\Omega $ slightly larger than $2$ in panels (B2) and (B3) is
readily explained by Eq. (\ref{2shifted}).

\begin{figure}[tbp]
\centering{\includegraphics[scale=0.4]{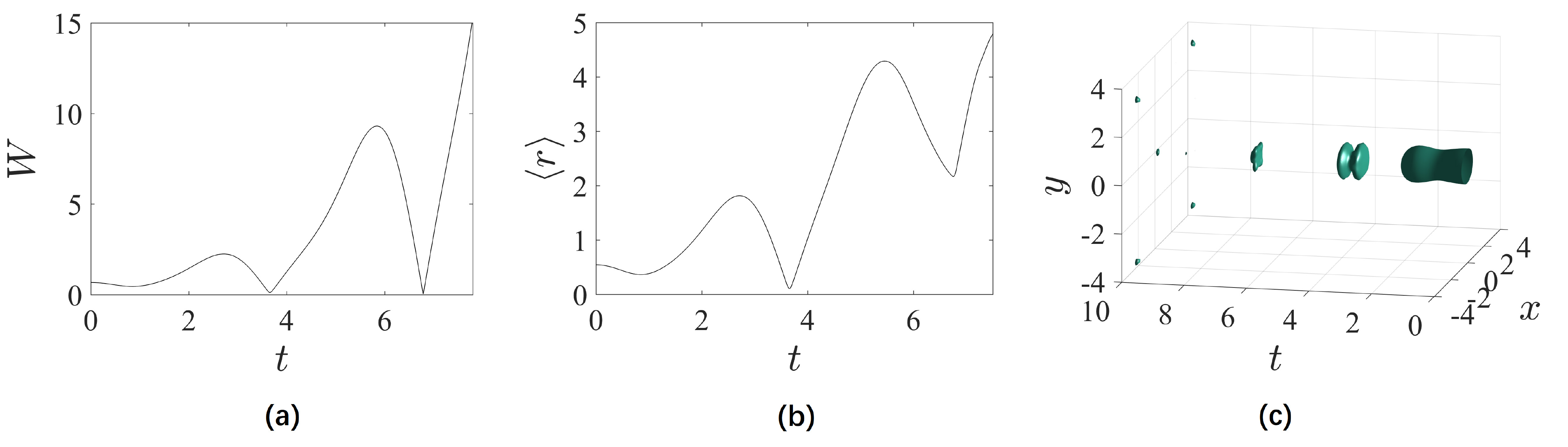}}
\caption{A typical example of the evolution of an unstable FS under the
action of TEM with parameters $(g,\protect\kappa _{\mathrm{ac}},\protect%
\omega )=(5,2,2)$. Results of simulations of the VA-produced Ermakov
equation (\protect\ref{eq-W}) and of GPE (\protect\ref{eq-model2}) are
displayed, respectively, in panels (a) and (b,c). In panel (a) it is seen
that, after several oscillations of the width with an increasing amplitude,
fast expansion of the wave function commences. Panels (b) and (c) display
essentially the same outcome of the evolution, by means of the time
dependence of $\left\langle {r}\right\rangle (t)$ and density profile.}
\label{fig03}
\end{figure}

A typical example of the PR-driven instability of the ac-driven FS is
displayed in Fig. \ref{fig03}. This example is produced for $\omega =2$,
which directly corresponds to the fundamental PR, as given by Eq. (\ref%
{param-res}) with $n=0$. It is seen that both full simulations of the
underlying GPE (\ref{eq-model2}) and the corresponding numerical solution of
the variational equation (\ref{eq-W}) lead to decay of the trapped FS, which
takes place after several oscillations with an increasing amplitude, in the
interval of time which is identified, approximately, as $0<t<7.5$. At $%
t\approx 7.5$, the expanding FS hits the region where the above-mentioned
edge absorber is installed. The GPE simulations can be extended to larger
times, but the results are then essentially affected by the loss inflicted
by the absorber. In any case, the onset of the PR-induced instability is
adequately revealed by the GPE\ simulations at the stage which is not
affected by the absorber. The same is true as concerns the onset of the
instability at other values of parameters.

\begin{figure}[tbp]
\centering{\includegraphics[scale=0.6]{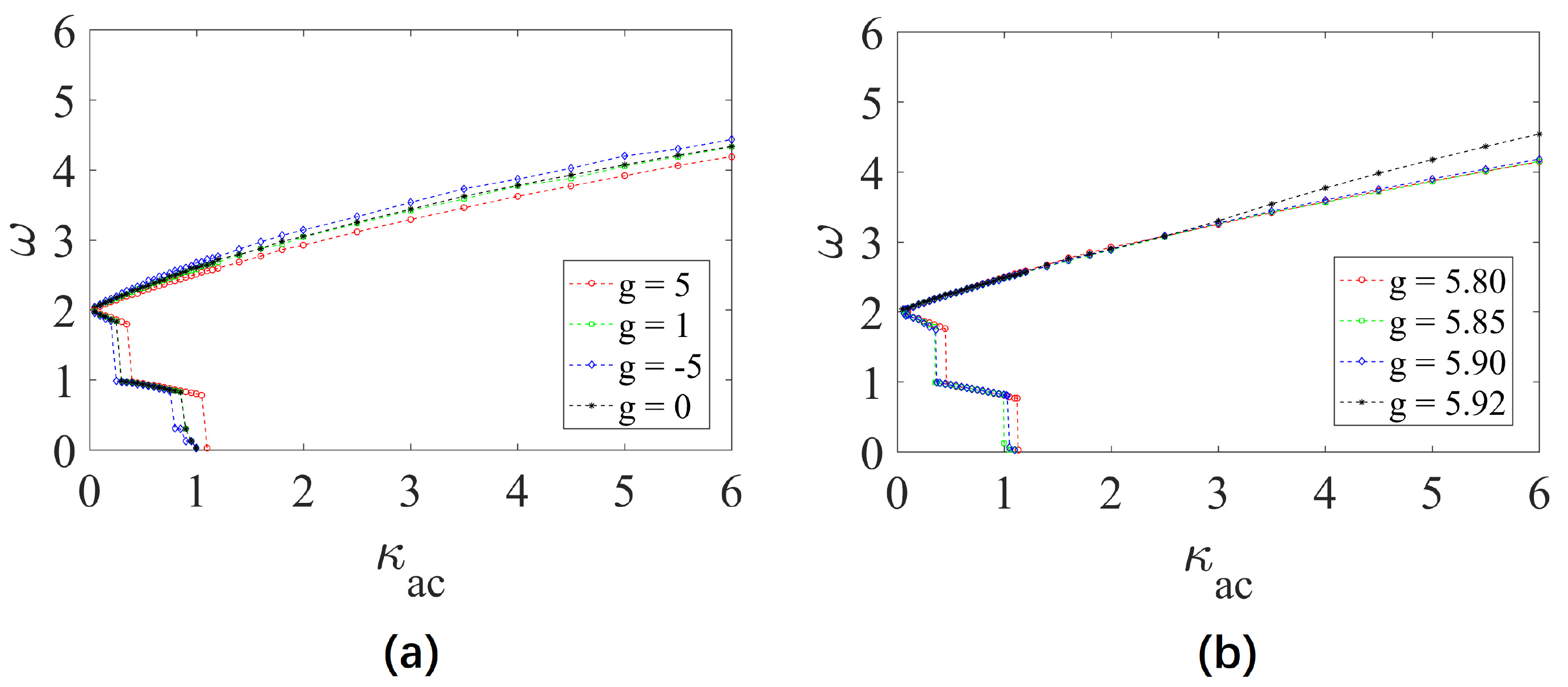}}
\caption{The stability diagram in the $\left( \protect\kappa _{\mathrm{ac}},%
\protect\omega \right) $ plane for the ac-driven FS, as produced by
systematic simulations of GPE (\protect\ref{eq-model2}). The instability
takes place between the top and bottom boundaries. (a) A set of boundaries
for $g=\pm 5$ and $g=0,1$, which includes both the self-attractive and
repulsive signs of the nonlinearity, as well as the linear system ($g=0$).
(b) A set of the stability diagrams for values of $g$ close to the threshold
of the critical collapse, at which the instability is still determined by
the PR (not by the proximity to the collapse).}
\label{fig04}
\end{figure}

Results produced by the systematic simulations of GPE (\ref{eq-model2}) at
different values of parameters $\omega $, $\kappa _{\mathrm{ac}}$, and $g$
(including both $g>0$ and $g<0$, i.e., the self-attractive and repulsive
nonlinearity) are collected in the form of the stability diagrams shown in
Fig. \ref{fig04}. The ac-driven FS is stable beneath the boundaries shown in
the left bottom corner of the plots and above boundaries in their top parts.
First, point $\omega =2$ at which the instability area emerges is explained
by the fundamental PR, which is predicted by Eq. (\ref{param-res}) with $n=0$%
. Further, a relatively short segment of the bottom boundary at $\omega
\approx 1$ is explained as a manifestation of the next-order PR
corresponding to $n=1$ in Eq. (\ref{param-res}). The instability does not
occur at very large values of $\omega $. This case can be considered by
means of the above-mentioned averaging method, which does not demonstrate
any source of instability, cf. Ref. \cite{Galimzyanov}.

Figure \ref{fig04} shows that the dependence of the stability boundaries on
the strength and sign of the nonlinearity [in particular, in panel (a),
which shows the results for $g=\pm 5$] is very weak, including values of $g$
in panel (b) which are close to the collapse threshold, cf. Eq. (\ref{gg}).
In this connection, it is relevant to recall that, in the framework of the
nonlinear Ermakov equation (\ref{eq-W}), the boundary of the PR-induced
instability indeed does not depend on the nonlinearity strength, being
identical to that in the linear Mathieu equation (\ref{Mathieu}). Actually,
an essential result demonstrated by Fig. \ref{fig04} is that the full GPE (%
\ref{eq-model2}), for which the VA equations (\ref{eq-W}) and (\ref{Mathieu}%
) are only an approximation, produces a visible but very weak dependence of
the boundary on $g$, i.e., the approximation is quite accurate, in this
sense.

Note that the stability diagrams displayed in Fig. \ref{fig04} include not
only values $\kappa _{\mathrm{ac}}>1$, for which the sign of the quadratic
potential in Eq. (\ref{eq-model2}) periodically flips, but also $\kappa _{%
\mathrm{ac}}<1$, for which the potential always keeps the trapping sign.
Accordingly, the system is more robust in the latter case (usually, only
this case is considered in the framework of the Mathieu equation \ref%
{Mathieu}). Indeed, the bottom stability area in all panels of Fig. (\ref%
{fig04}) exists solely at $\kappa _{\mathrm{ac}}<1$.

\begin{figure}[tbp]
\centering{\includegraphics[scale=0.4]{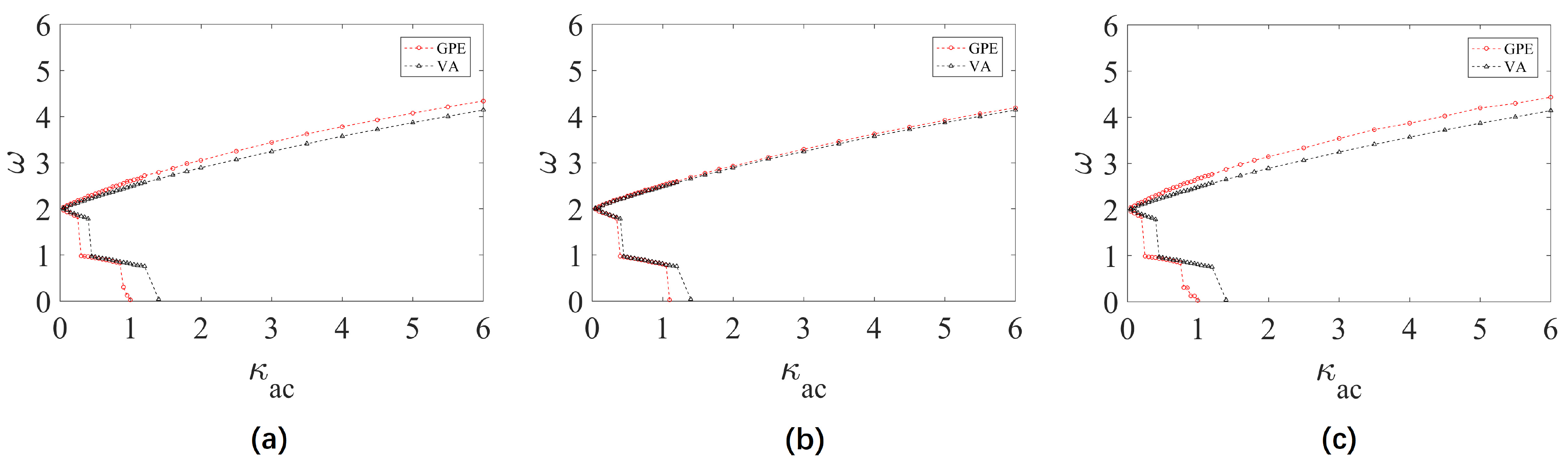}}
\caption{Stability boundaries in the plane of $(\protect\kappa _{\mathrm{ac}%
},\protect\omega )$, as produced by the simulations of the GPE in the form
of Eqs. (\protect\ref{eq-model2}) and (\protect\ref{eq-kappa}), and by the
numerical solution of the VA-predicted Ermakov equation (\protect\ref{eq-W}%
). The nonlinearity coefficient is $g=0$ in (a) (i.e., the system is
linear), $g=5$ in (b), and $g=-5$ in (c). }
\label{fig05}
\end{figure}

The stability boundaries, as produced by the systematic simulations of the
GPE (\ref{eq-model2}) and by the numerical solution of the VA-produced
Ermakov equation (\ref{eq-W}), are compared in Fig. \ref{fig05}, by
juxtaposing them in the plane of $\left( \kappa _{\mathrm{ac}},\omega
\right) $ for $g=0$, $1$, and $\pm 5$. It is seen that the VA always
provides a reasonable agreement with the full GPE, and becomes very accurate
for larger values of $g$.

\subsection{The boundary of the critical collapse}

\begin{figure}[tbp]
\centering{\includegraphics[scale=0.4]{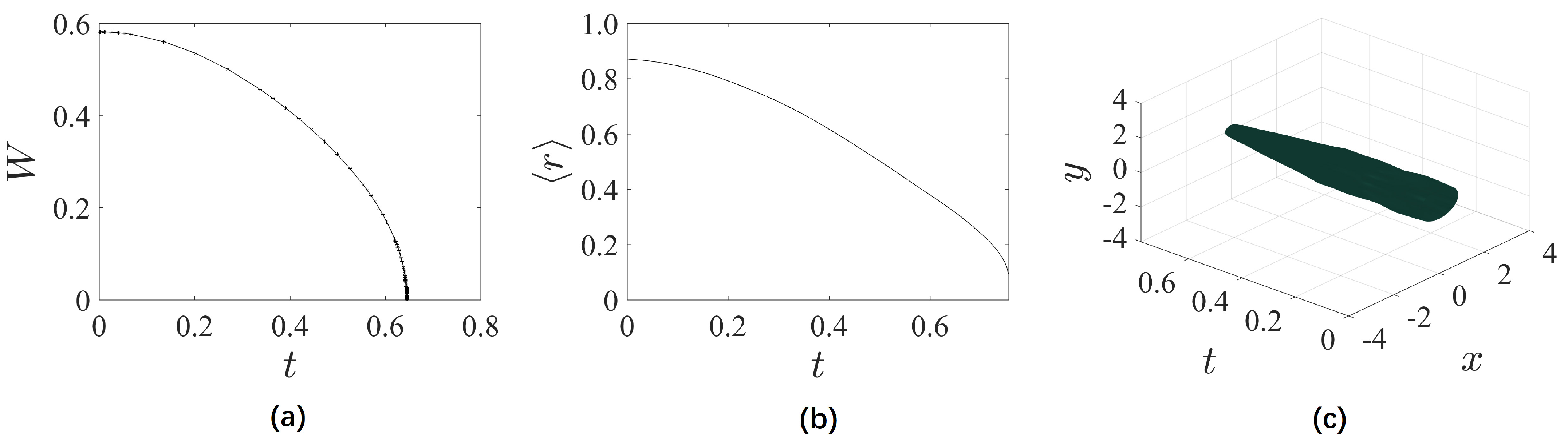}}
\caption{A typical example of the evolution of a collapsing FS obtained for
parameters $(g,\protect\kappa _{\mathrm{ac}},\protect\omega )=(7,2,4)$. (a)
The shrinking width, $W(t)$, as predicted by the VA, i.e., by the Ermakov
equation (\protect\ref{eq-W}). (b,c) The shrinkage of the FS as produced by
the simulations of GPE (\protect\ref{eq-model2}). The plots in panels (a)
and (b,c) are cut, respectively, at times indicated in Eq. (\protect\ref{tt}%
).}
\label{fig06}
\end{figure}

A typical example of the solution which quickly develops the collapse at $%
g=7 $, that definitely exceeds the largest value admitting the stability of
the FS, is displayed in Fig. \ref{fig06}. The blowup of the solution is
obvious. In particular, its VA-predicted width shrinks to zero at the
collapse moment, cf. the VA-predicted solution for the collapse in the
absence of the HO potential, given by Eqs. (\ref{W(z)}) and (\ref{tcoll}).
In this case, the collapse takes place after the evolution time%
\begin{equation}
t_{\text{\textrm{collapse}}}^{\mathrm{(VA)}}\approx 0.64;~t_{\text{\textrm{%
collapse}}}^{\mathrm{(numer)}}\approx 0.76  \label{tt}
\end{equation}%
in the framework of the VA or GPE simulations, respectively, which is
essentially smaller than the TEM period, $2\pi /\omega \approx \allowbreak
1.57$, i.e., the time modulation of the trapping potential does not
essentially affect the onset of the collapse.

\begin{figure}[tbp]
\centering{\includegraphics[scale=0.6]{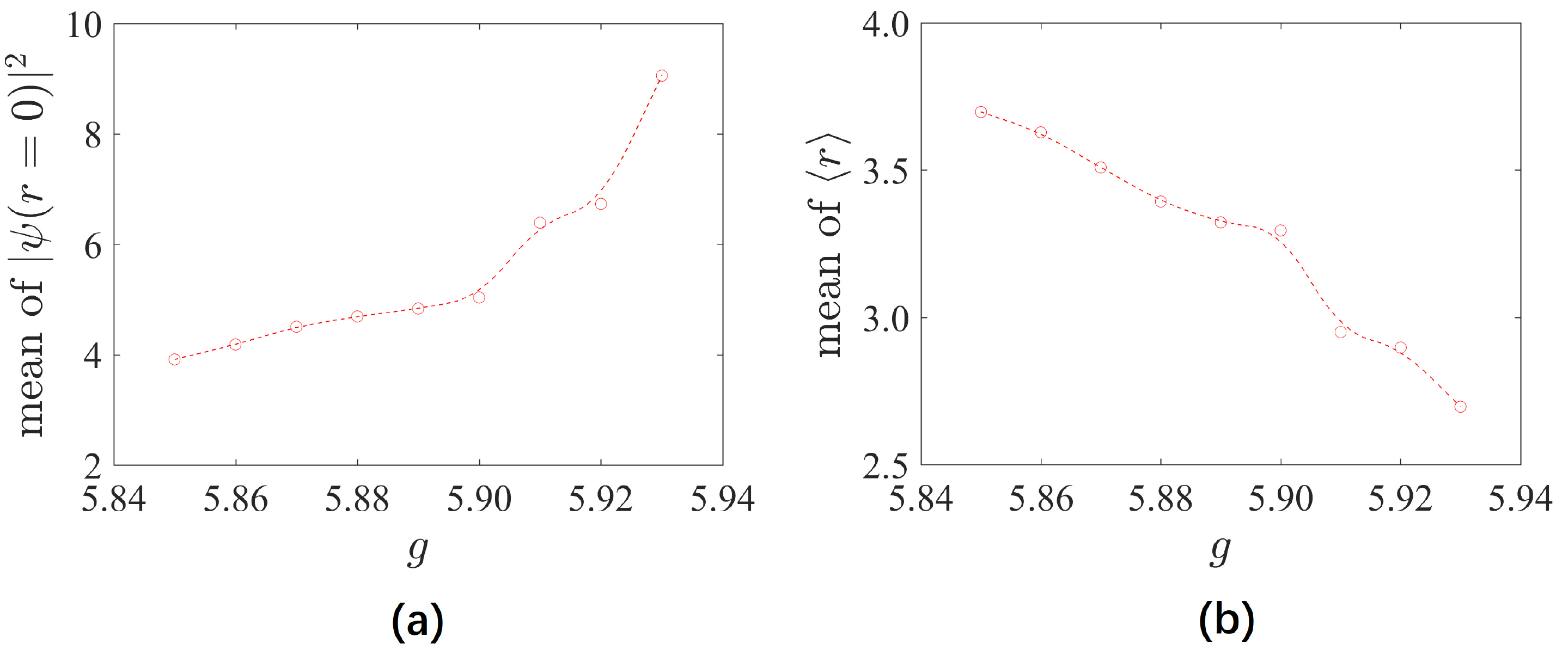}}
\caption{Dependences of mean (time-average) values of the density at the
central point (a) and radial size (b), produced by simulations of Eq. (%
\protect\ref{eq-model2}) and (\protect\ref{eq-kappa}) with $\protect\kappa _{%
\mathrm{ac}}=2$ and $\protect\omega =4$, on the self-attraction strength, $g$%
. The dependences terminate at the collapse point, $g_{\mathrm{c}}\approx
5.938$. This value exceeds the standard one, $g_{\mathrm{c}}\approx 5.85$,
corresponding to the usual TSs (Townes solitons), see Eq. (\protect\ref{gg}%
). }
\label{fig07}
\end{figure}

Because, as mentioned above, the difference between the numerically exact
and VA-predicted values of the self-attraction strength at the point of the
onset of the critical collapse, in the absence of TEM, is conspicuous, $%
\left( g_{\mathrm{c}}^{\mathrm{(VA)}}-g_{\mathrm{c}}^{\mathrm{(num)}}\right)
/g_{\mathrm{c}}^{\mathrm{(VA)}}\simeq 7\%$, as per Eq. (\ref{gg}), the VA is
not appropriate for accurate identification of the shift of $g_{\mathrm{c}}$
under the action of TEM. This was done by means of systematic simulations of
GPE (\ref{eq-model2}) with the HO strength taken according to Eq. (\ref%
{eq-kappa}). An example of the implementation of this approach is displayed
in Fig. \ref{fig07}. It shows that the time-average (mean) values of the
peak density and radial size of the ac-driven SF monotonously grow and
decrease, respectively, with the increase of $g$, passing the value $g_{%
\mathrm{c}}^{\mathrm{(num)}}\approx 5.85$, which corresponds to the usual TS
[see Eq. (\ref{gg})], and attaining the critical point at $g_{\mathrm{c}%
}\approx 5.938$, at which the solution suffers the blowup.

\begin{figure}[tbp]
\centering{\includegraphics[scale=0.58]{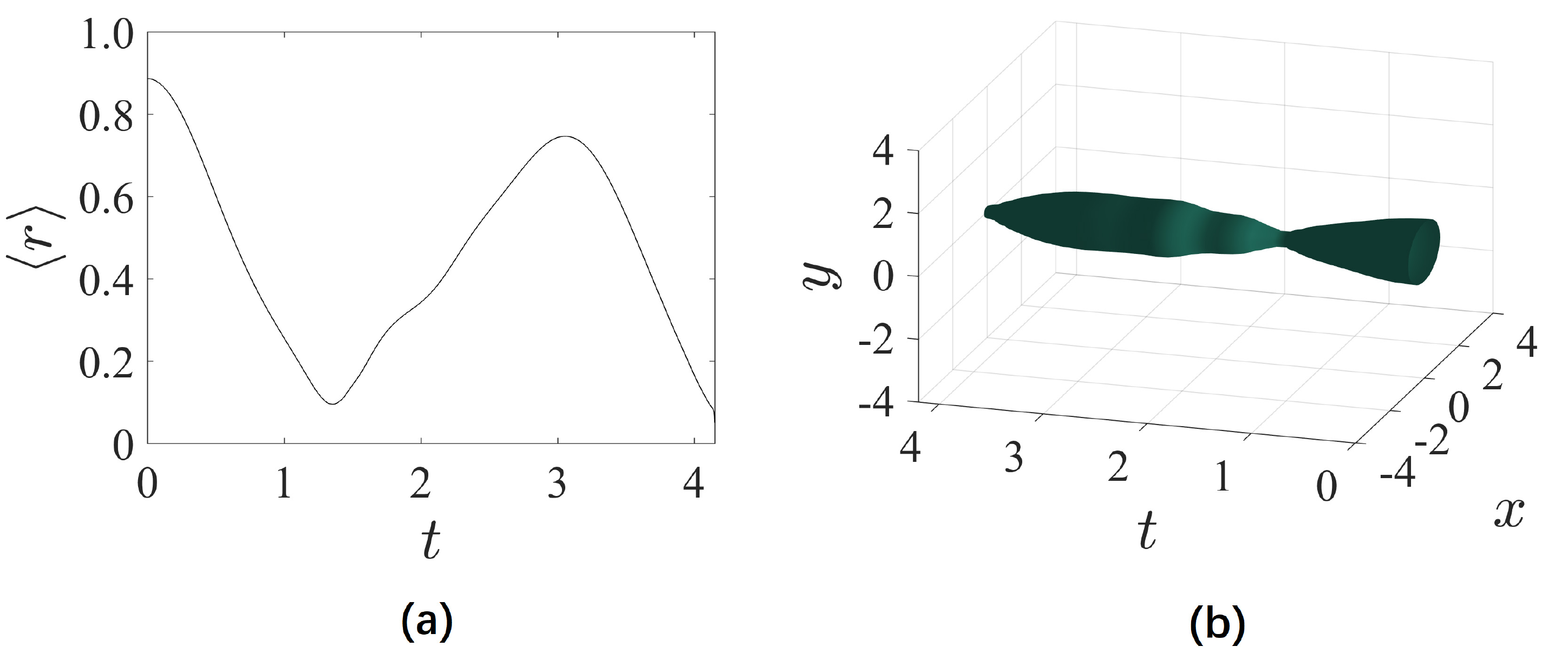}}
\caption{Panels (a) and (b) display the same as in Figs. \protect\ref{fig06}%
(b,c), but for parameters $(g,\protect\kappa _{\mathrm{ac}},\protect\omega %
)=(5.938,2,4)$. }
\label{fig08}
\end{figure}

The development of the collapse at a point which is almost exactly
tantamount to the critical one is displayed in Fig. \ref{fig08}. Due to the
action of the TEM, the collapse takes place after a few cycles of
compression and expansion of the FS.

\begin{figure}[tbp]
\centering{\includegraphics[scale=0.4]{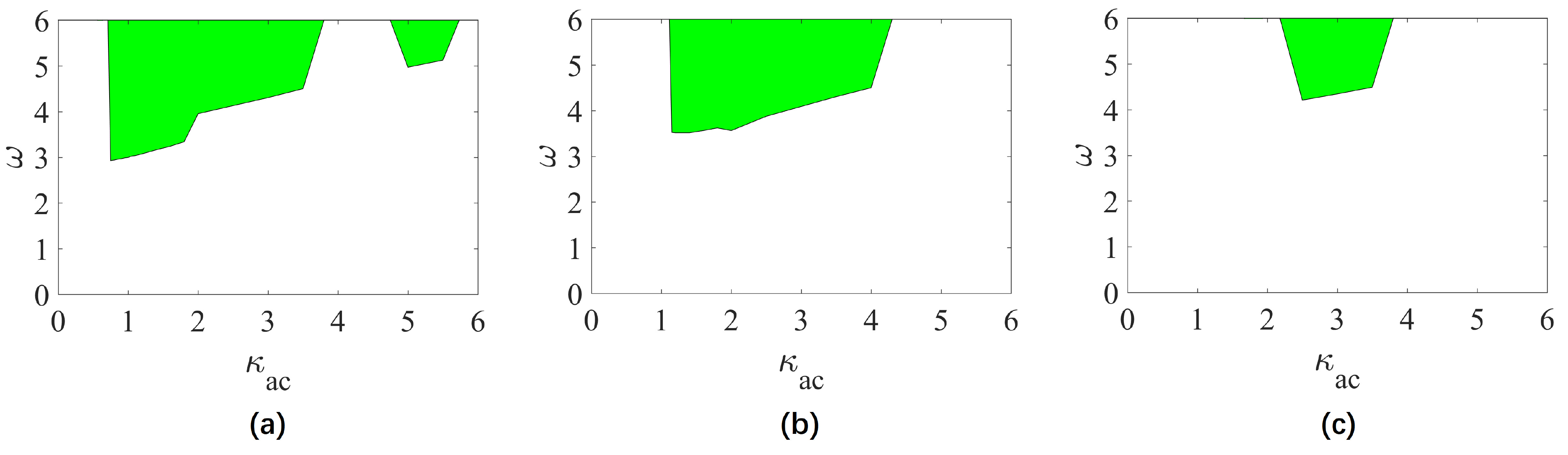}}
\caption{Stability diagrams in the plane of $(\protect\kappa _{\mathrm{ac}},%
\protect\omega )$ at values of the self-attraction strength, $g$, exceeding
the usual critical value, $g_{\mathrm{c}}^{\mathrm{(num)}}\approx 5.85$, see
Eq. (\protect\ref{gg}). Stable FSs are produced by simulations of Eqs. (%
\protect\ref{eq-model2}) and (\protect\ref{eq-kappa}) in green areas. Fixed
values of $g$ are $5.933$ in (a), $5.935$ in (b), and $5.937$ in (c).}
\label{fig09}
\end{figure}

In Fig. \ref{fig09} we plot the boundary of the onset of the collapse in the
plane of $\left( \kappa _{\mathrm{ac}},\omega \right) $, for three value of
the self-attraction strength $g$ exceeding the usual critical value $\approx
5.85$ [see Eq. (\ref{gg})], as produced by the systematic simulations of
Eqs. (\ref{eq-model2}) and (\ref{eq-kappa}). A characteristic difference
between the boundary determined by the onset of the critical collapse, and
the boundary of the PR-induced instability plotted in the same parametric
plane at $g<g_{\mathrm{c}}$ (cf. Figs. \ref{fig04} and \ref{fig05}), is that
the collapse boundary shrinks to one or two segments, no stable FS existing
outside of them.

\begin{figure}[tbp]
\centering{\includegraphics[scale=0.6]{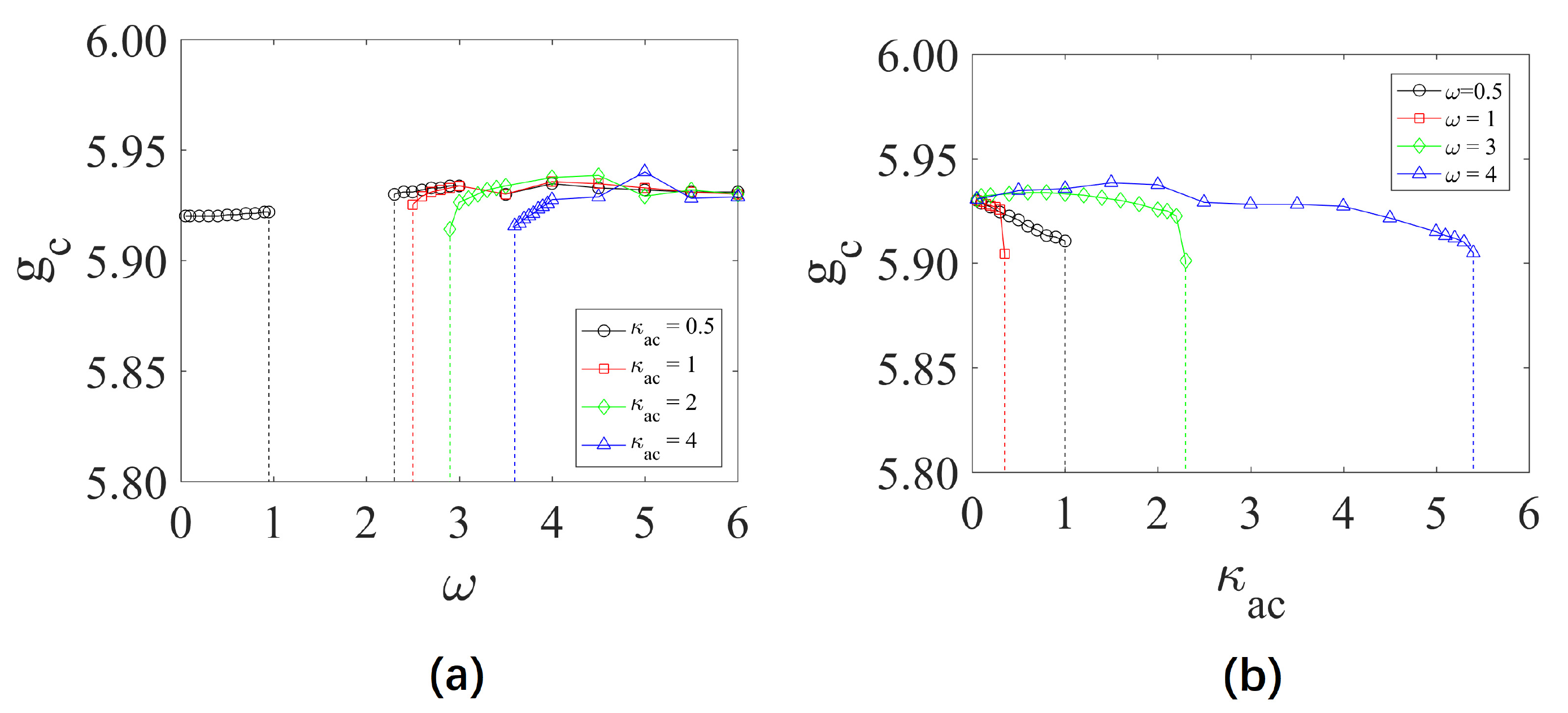}}
\caption{(a) The critical value, $g_{\mathrm{c}}$, of the self-attraction
strength, $g$, at which the trapped FS is destroyed by the blowup, vs. the
driving frequency, $\protect\omega $, at several fixed values of ac-drive's
amplitude, $\protect\kappa _{\mathrm{ac}}$. (b) The same for $g_{\mathrm{c}}$
as a function of $\protect\kappa _{\mathrm{ac}}$ at fixed values of $\protect%
\omega $. Gaps in which the dependences are not displayed are regions in
which $g_{\mathrm{c}}$ falls to essentially lower values.}
\label{fig10}
\end{figure}

Finally, the most important characteristic of the partial stabilization of
the FS modes by TEM at
\begin{equation}
g>g_{\mathrm{c}}^{\mathrm{(num)}}\approx 5.85  \label{g>g}
\end{equation}%
is presented by Fig. \ref{fig10}, which shows the critical value $g_{\mathrm{%
c}}$ of the self-attraction strength, at which the solutions suffer the
blowup, vs. $\omega $ and $\kappa _{\mathrm{ac}}$. Similar to the situation
observed in Fig. \ref{fig09}, the dependences feature gaps, in which the
blowup commences at essentially lower values of $g$. The largest value of $%
g_{\mathrm{c}}$ produced by the systematic simulations is $\left( g_{\mathrm{%
c}}\right) _{\max }\approx 5.938$.

Note that the results summarized in Figs. \ref{fig09} and \ref{fig10} are
obtained for both cases of $\kappa _{\mathrm{ac}}<1$ and $\kappa _{\mathrm{ac%
}}>1$. Generally, the latter case is more favorable for the expansion of the
stability region into the nontrivial area (\ref{g>g}), unless $\omega $ is
too small (in particular, Fig. \ref{fig09} shows no stability regions at $%
\kappa _{\mathrm{ac}}<1$). These trends can be understood, as Eq. (\ref%
{eq-kappa}) with $\kappa _{\mathrm{ac}}>1$ implies the periodic switching
from the trapping (HO) potential to the expulsive (anti-HO) one. Naturally,
the anti-HO potential tends to arrest the evolution towards the blowup at $%
r=0$. This feature can be demonstrated in an approximate form by the fact
that Eq. (\ref{eq-W}) (in which $N=1$ is set, as above), with $\kappa _{%
\mathrm{dc}}+\kappa _{\mathrm{ac}}\cos \left( \omega t\right) $ replaced by
a constant \emph{negative} value of $\kappa $, admits an FP solution in the
case when $g$ \emph{exceeds} the respective critical value, $g_{\mathrm{c}}^{%
\mathrm{(VA)}}=2\pi $ [see Eq. (\ref{gg})]:
\begin{equation}
W_{\mathrm{FP}}=\left( -\frac{1}{\kappa }\left( \frac{g}{2\pi }-1\right)
\right) ^{1/4},  \label{W0}
\end{equation}%
cf. stationary solution (\ref{eq-W0stab}). Of course, the FP given by Eq. (%
\ref{W0}) is, by itself, unstable, unlike its counterpart (\ref{eq-W0stab}),
but its appearance helps to understand how TEM makes the FS more robust
against the collapse. On the other hand, this mechanism is not efficient for
low ac-driving frequencies $\omega $, as the corresponding time $\sim \pi
/\omega $, needed for the switch between the HO\ and anti-HO potentials, may
be larger than the collapse time, which is approximately given by Eq. (\ref%
{tcoll}).

\begin{figure}[tbp]
\centering{\includegraphics[scale=0.4]{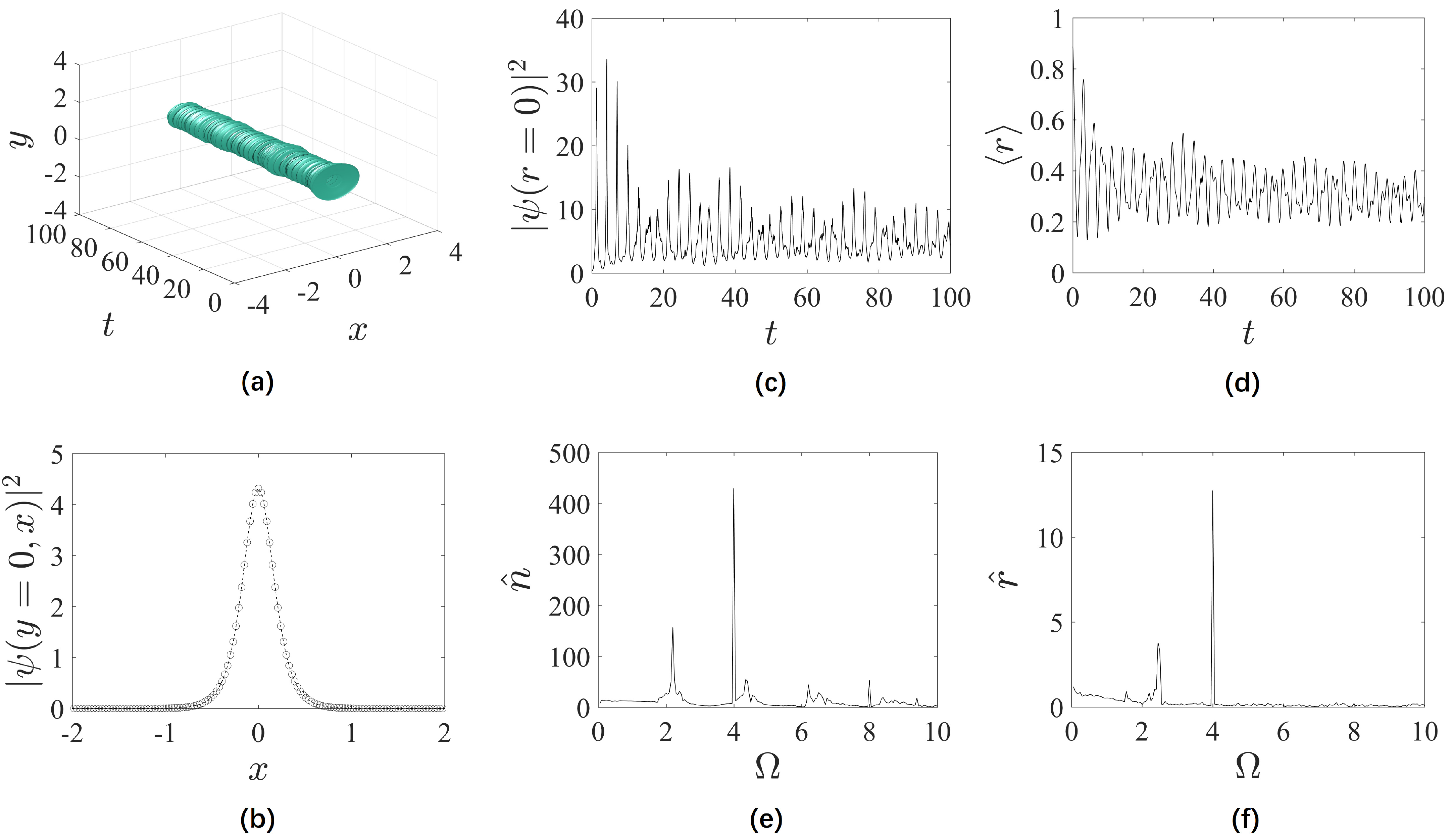}}
\caption{A typical stable FS found in region (\protect\ref{g>g}), at
parameters $(g,\protect\kappa _{\mathrm{dc}},\protect\kappa _{\mathrm{ac}},%
\protect\omega )=(5.90,1,2,4)$. The solution is produced by simulations of
Eq. (\protect\ref{eq-model2}) with input (\protect\ref{input}). (a) The
spatiotemporal density profile. (b) The cross section of the spatial
profile, $|\protect\psi (y=0,x)|^{2}$ at $t=100$. (c) and (d): The evolution
of the peak power, $\left\vert \protect\psi \left( x=y=0,t\right)
\right\vert ^{2}$, and effective radius (\protect\ref{r}). (e) and (f):
Spectra of the Fourier transform of the same variables, defined as per Eq. (%
\protect\ref{Fourier}).}
\label{fig11}
\end{figure}

\begin{figure}[tbp]
\centering{\includegraphics[scale=0.3]{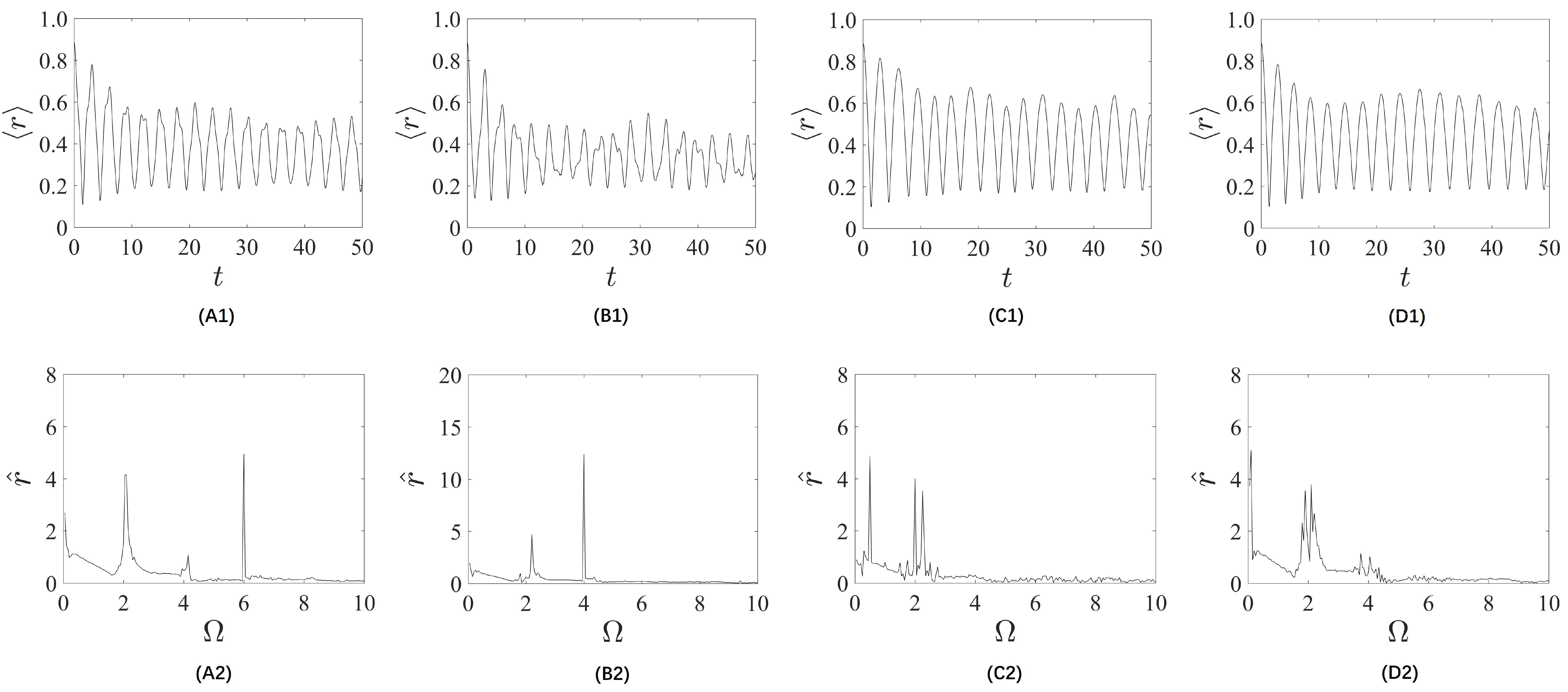}}
\caption{The same as in Figs. \protect\ref{fig02}(A1-D1) and (A2-D2), but
for $(g,\protect\kappa _{\mathrm{dc}})=(5.90,1)$. Note that the solution
displayed in panels (B1) and (B2) is the same as shown in Fig. \protect\ref%
{fig11}. It is included here for the comparison's sake.}
\label{fig12}
\end{figure}

A typical example of an FS found in the nontrivial stability area (\ref{g>g}%
), \textit{viz}., at $g=5.90$, is displayed in Fig. \ref{fig11}. The
dynamical structure of such states is generally similar to that presented in
Fig. \ref{fig02} for $g=1$. Characteristic examples, for the same cases of
the low- and high-frequency ac drive as those in Fig. \ref{fig02}, are
displayed in Fig. \ref{fig12}. Like in Fig. \ref{fig02}, the spectrum
features two main peaks, at $\Omega =\omega $ and $\Omega \simeq 2$. The
latter one, predicted by Eqs. (\ref{2}) and (\ref{2shifted}), features a
split shape in the case of the low-frequency drive. On the other hand, a
difference is that the time dependences of the peak density and radial size,
as well as the respective spectra, demonstrate robust but irregular
oscillations in Fig. \ref{fig12}, unlike the quasi-regular dynamical regimes
revealed by Fig. \ref{fig02}.
{\LARGE \ }

\begin{figure}[tbp]
\centering{\includegraphics[scale=0.3]{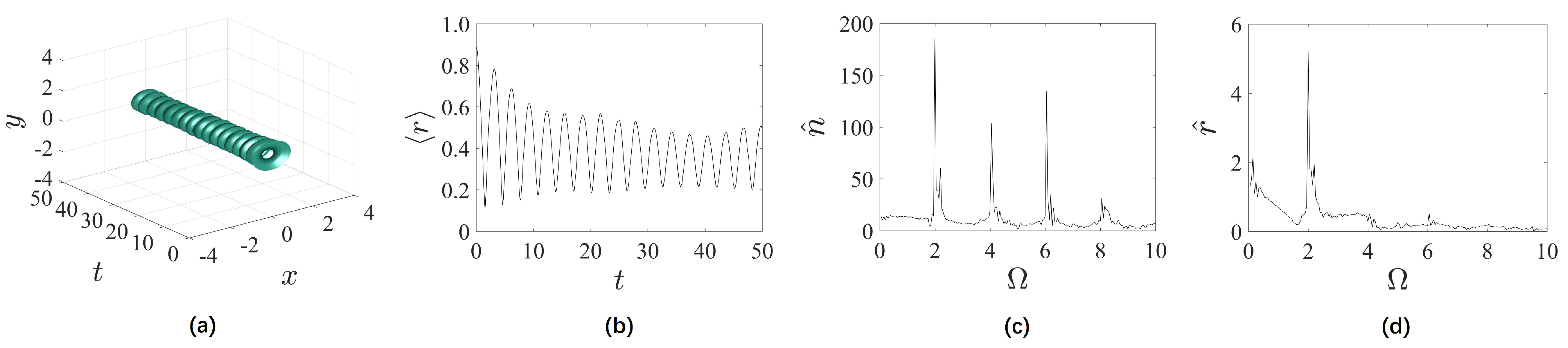}}
\caption{The stable evolution of a breather in the absence of the ac drive
in Eq. (\protect\ref{eq-model2}), with parameters $(g,\protect\kappa _{%
\mathrm{dc}},\protect\kappa _{\mathrm{ac}})=(5.90,1,0)$. (a) The evolution
of the density profile. (b) The evolution of the monopole moment (radial
size). (c) and (d): Fourier transform of the peak density, $\hat{n}(\Omega )$%
, and radial size, $\hat{r}(\Omega )$, calculated as per Eq. (\protect\ref%
{Fourier}).}
\label{fig13}
\end{figure}

A noteworthy peculiarity of Figs. \ref{fig10}(a) and (b) is that the
nontrivial stability region (\ref{g>g}) exists, up to $g\approx 5.93$, even
in the limit cases of $\omega =0$ and $\kappa _{\mathrm{ac}}=0$, when the ac
drive is not present. In this case, simulations of Eq. (\ref{eq-model2}),
with the input taken as a stationary FS numerically found at $g\leq g_{%
\mathrm{c}}^{\mathrm{(num)}}\approx 5.85$, lead to the collapse, as it might
be expected. However, the input (\ref{input}) produces oscillatory states
(breathers) which, unlike the collapsing quasi-stationary states, may indeed
keep their dynamical stability up to $g\approx 5.93$. An example of such a
dynamical regime, produced by the simulations of Eq. (\ref{eq-model2}) with $%
g=5.90$ and $\kappa _{\mathrm{ac}}=0$, is presented in Fig. \ref{fig13}. In
particular, the power spectrum observed in Fig. \ref{fig13}(c) is typical
for oscillations in autonomous nonlinear dynamical systems, being different
from that for the ac-driven system, cf. Fig. \ref{fig11}(e).

\section{Conclusion}

The aim of this work is to elaborate a scheme of TEM (trapping-expulsion
management), which helps to stabilize 2D FS (fundamental-state) modes under
the action of the cubic self-attraction, that gives rise to the critical
collapse in the 2D space, thus making the usual TSs (Townes solitons)
completely unstable. The TEM scheme works by applying the quadratic
potential with the periodically flipping sign, so that it switches between
the trapping HO (harmonic-oscillator) and expulsive anti-HO forms. The TEM
scenario can be realized in nonlinear optics and in BEC. The analysis of the
FS dynamics under the action of TEM is performed by means of systematics
simulations, in the combination with the VA (variational approximation). The
VA reduces the FS dynamics to an equation of the Ermakov type. Stability
boundaries for the FS trapped in the periodically switching potential have
been identified, as functions of strength $g$ of the cubic self-attraction,
and amplitude and frequency of the ac (time-periodic) part of the
potential's strength. Below the standard (Townes) collapse threshold, which
means $g<g_{\mathrm{c}}^{\mathrm{(num)}}\approx 5.85$ in the notation
adopted here, the stability area is bounded by the onset of the fundamental
or higher-order PR (parametric resonance). This boundary is well
approximated by the VA, including the system with the self-repulsive
nonlinearity, $g<0$, and the linear one ($g=0$). At $g>g_{\mathrm{c}}$, the
collapse boundary is identified for FSs by means of systematic simulations
of the underlying GPE (Gross-Pitaevskii equation). The largest value of the
self-attraction strength, which admits the stability against the critical
collapse in the system, is $g\approx 5.938$, exceeding the standard one, $g_{%
\mathrm{c}}^{\mathrm{(num)}}\approx 5.85$, by $\approx 1.5\%$. This
increase, although relatively small, is a significant result, as the
standard collapse threshold is usually strictly fixed. The extension of the
stability region above $g_{\mathrm{c}}^{\mathrm{(num)}}\approx 5.85$ takes
place also in the absence of the TEM. It is explained by the fact that a
breather, in the form of an oscillatory FS, may keep its stability against
the collapse at values of $g$ which are somewhat larger than $5.85$.

The obtained results can be verified in experiments, and, as concerns the
implementation in optics, they may find applications to the design of
photonic devices based on waveguide-antiwaveguide schemes.

A natural direction for the continuation of the work is the development of
the analysis for 2D states with intrinsic vorticity, which will be reported
elsewhere. Another possibility is the consideration of TEM in the 1D system
with the quintic self-attraction, which gives rise to the TSs and critical
collapse in 1D \cite{quintic1,quintic2}. It may also be interesting to
consider the 2D model which includes, as an additional stabilizing
ingredient, a permanent anharmonic (quartic) term in the trapping potential.
Such a term was considered in the context of BEC in various settings \cite%
{quartic,quartic0,quartic1,quartic2}.

\section{Acknowledgments}

This work was supported by NNSFC (China) through grant Nos.
1187411211905032, Natural Science Foundation of Guangdong province through
grant No. 2021A1515010214, the Key Research Projects of General Colleges in
Guangdong Province through grant No. 2019KZDXM001, the Research Fund of
Guangdong-Hong Kong-Macao Joint Laboratory for Intelligent Micro-Nano
Optoelectronic Technology through grant No. 2020B1212030010, and by the
Israel Science Foundation through grant No. 1286/17.

\end{document}